# X-ray Photons in the CO 2-1 'Lacuna' of NGC 2110

*G. Fabbiano[1], A. Paggi[2,3], M. Elvis[1]*

*1. Center for Astrophysics | Harvard & Smithsonian, 60 Garden St. Cambridge MA 02138, USA*
*2. INAF-Osservatorio Astrofisico di Torino, via Osservatorio 20, 10025 Pino Torinese, Italy*
*3. Istituto Nazionale di Fisica Nucleare, Sezione di Torino, via Pietro Giuria 1, 10125 Torino, Italy*

## Abstract

A recent ALMA study of the Seyfert 2 Active Galactic Nucleus (AGN) NGC 2110 by Rosario et al. (2019) has reported a remarkable lack of CO 2-1 emission from the circumnuclear region, where optical lines and $H_2$ emission are observed, leading to the suggestion of excitation of the molecular clouds by the AGN. Since interaction with X-ray photons could be the cause of this excitation, we have searched the archival *Chandra* data for corroborating evidence. We report the discovery of an extra-nuclear ~1'' (~170 pc) feature in the soft (<1.0 keV) *Chandra* data. This feature is elongated to the north of the nucleus and its shape matches well that of the optical lines and $H_2$ emission in this region, which is devoid of CO 2-1 emission. The *Chandra* image completes the emerging picture of a multi-phase circumnuclear medium excited by the X-rays from the AGN, with dense warm molecular clouds emitting in $H_2$ but depleted of CO 2-1 line emission.

## 1. Introduction

NGC 2110 is a well-studied, nearby (~35 Mpc, 170 pc /arcsec; NED) Seyfert 2 galaxy, associated with a ~4'' size double lobed plus core radio source (Ulvestad & Wilson 1983). The AGN is variable in the X-rays (Mushotzky 1982), radio (Mundell et al. 2009), and Hα line (Moran et al. 2007). It has a complex absorption system (Rivers et al. 2014), consisting of a steady component with $N_H=3\times10^{22}$ cm$^{-2}$ plus a variable absorber associated with the broad line region. Based on the $M_{BH}$-σ relation, Moran et al. (2007) estimate the mass of the nuclear black hole to be $2\times10^8$ M$_\odot$. Extended optical and IR line emission with complex kinematics are associated with this AGN (Wilson et al. 1985; Fischer et al. 2013; Schnorr- Müller et al. 2014; Diniz et al. 2015).

A recent paper by Rosario et al. (2019) reports a high resolution ALMA study of the nuclear region of NGC 2110. Comparing these high resolution data with optical line emission from HST and SINFONI, these authors report the presence of a central ~1'' N-S CO cavity in the region of most intense optical line emission, and conclude that the CO 2-1 emission is suppressed because of the interaction of the AGN with the molecular gas. Warming up of the near-nucleus molecular clouds by the AGN is also suggested by the H$_2$ image of the nuclear region (Diniz et al 2015). Are the vectors of the AGN feedback the nuclear radio jets (via shocks), or is photo-excitation from X-rays occurring (e.g., Riffel et al. 2008; Wang et al. 2011 )?

To investigate this question, we have taken a new look at the *Chandra* ACIS observation of NGC 2110, with the aim of mapping the X-ray emission at high resolution to compare it with the optical line emission and the CO cavity. Extended emission was first reported in soft X-rays, peaked ~4'' north of the nucleus, based on ROSAT HRI observations (Weaver et al. 1995). With *Chandra* ACIS, Evans et al. (2006) also reported extended emission associated with NGC 2110. This emission is elongated in the same N-S direction as the VLA radio, 5 GHz, emission and the HST [OIII] line emission. However, the details of the extended X-ray emission, especially at smaller radii (<2'') were not well defined, mostly due to the presence of a strong nuclear source that reduces contrast of the extended emission at these scales, and which saturates the ACIS read-out leading to pile-up (see the Chandra Proposers' Observatory Guide, POG, Sections 6.13.1 and 6.16)[1].

With $N_H$=3×10$^{22}$ cm$^{-2}$ or larger (Rivers et al 2014), the nuclear emission should be depressed by absorption at energies below 2 keV. By inspecting the *Chandra* archival data, we find that at energies below 1 keV the nucleus is sufficiently obscured so that the point source does not overpower the surrounding regions. Restricting ourselves to the soft energies (0.1-1 keV), we can then safely explore the extended emission as close to the nucleus as allowed by the angular resolution of *Chandra*. To correct for the ACIS under-sampling of the *Chandra* PSF and optimize the resolution of the data (Section 2), we use the now well-established sub-pixel analysis techniques (see e.g., Fabbiano et al. 2018b). The results (Section 3) show a remarkable similarity with the HST Hα+[N II] contours reported in Rosario et al. (2019), arguing for X-ray excitation of the molecular gas (Section 4).

## 2. Data Reduction and Analysis

We used the same ACIS-S observation (ObsID= 833; exposure time=45.7 ks; PI: Weaver) as used by Evans et al (2006). As described in that paper, the observation was performed in subarray mode (128 rows), with a 0.4s readout time that reduces the CCD pileup, although the pileup is still substantial in the nucleus (see Evans et al 2006). Pileup has the effect of depressing the count distribution in the pixels corresponding to the center of the PSF. If more than one photon arrives on a given pixel during the readout time frame, these counts are erroneously detected as a single higher energy photon. High count rate sources also present a 'read-out streak' that affects the quality of the image (see POG, Section 6.13.1).

---

[1] *Chandra* Proposers' Observatory Guide (POG), Rev. 21.0, December 2018
http://cxc.cfa.harvard.edu/proposer/POG/

ObsID 833 was acquired in April 22, 2000, when the soft response of ACIS-S was still excellent (POG, Section 6.5.1). By investigating images in different energy bands, we find that if we restrict our analysis to energies below 1.0 keV, while the shape of the point source surface brightness distribution is depressed at the center because of pileup, the nuclear emission has a low enough count rate that its wings do not overpower the surrounding emission. In this energy range, there is no sign of read-out streak. The count rate of the nuclear source in this energy band is ~0.007 counts s$^{-1}$. At higher energies, a read-out streak (roughly in the E-W direction) appears and the central source dominates the emission within ~2.5 arcseconds.

Level 2 event files were retrieved from the *Chandra* Data Archive and reduced with the CIAO (Fruscione et al. 2006) 4.10 software and the *Chandra* Calibration Data Base (CALDB) 4.8.1, adopting standard procedures.

The nucleus shows a significant ~ 15% pileup as measured by the CIAO tool PILEUP_MAP.

In the 0.1-1.0 keV band there is a very clear detection of emission extended in the N-S direction out to ~20'' (~3.4 kpc) from the nucleus (see also Evans et al 2006). To investigate the inner circumnuclear emission, we have followed the same procedures used for our high-resolution spatial analysis of the CT AGN ESO 428-G014 (see Section 2 of Fabbiano et al 2018b). Imaging analysis was performed without pixel randomization to take advantage of the telescope dithering in event positioning and with the subpixel event repositioning (SER) procedure (Li et al. 2003). We used a pixel size that was 1/2 and 1/16 of 0.492'', the native Chandra-ACIS detector pixel. This analysis is possible as the *Chandra* mirror PSF has a sharp peak on smaller scales, and the continuous 'dither' of *Chandra* accesses this information (POG). This technique has been validates in numerous papers (e.g. Harris et al. 2004; Siemiginowska et al. 2007; Wang et al. 2011a, b, c; Paggi et al. 2012; Fabbiano et al. 2018b).

We adaptively smoothed the soft band 0.1-1.0 keV image using the DMIMGADAPT tool, which preserve small-scale features while larger features are smoothed over a larger area, increasing the size of the convolution kernel until a minimum number of counts or a maximum kernel radius is reached. Details are given in Section 3 for each image. Any quantitative conclusions were reached by extracting counts from the unsmoothed data, to preserve and evaluate counts statistics. All X-ray images are displayed with north to the top and east to the left. The visual interface DS9[2] was used throughout to examine the data, produce the images displayed in this paper, and as interface to the CIAO image analysis.

We checked the position of the Chandra PSF artifact as obtained with the tool

---

[2] http://cxc.harvard.edu/ciao/workshop/aug11/ds9.pdf

MAKE_PSF_ASYMMETRY_REGION, and found that for the roll angle of ObsID 883 this is expected in the N-W direction from the nucleus, away from the detected feature (see Section 3).

## 3. Results

Figure 1 shows the large-scale ~1' (~10 kpc) extended emission. To produce this image, the data were binned in ½ ACIS pixel bins and adaptively smoothed with scales from ½ to 20 image pixels (~0''.25 – 10'') in 30 iterations, with a minimum of 9 counts under the Gaussian kernel. The extent is comparable to that of the 0.5-1.0 keV image of Evans et al (2006), but our choice of smoothing parameters suggests some structure in the surface brightness distribution to the south.

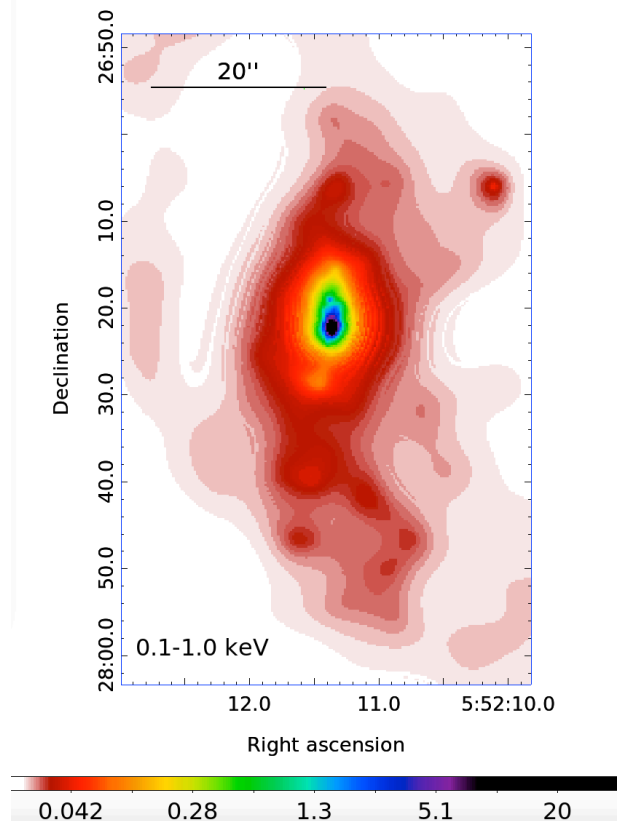

**Figure 1**. – Image of the extended circumnuclear emission of NGC 2110 in the 0.1-1.0 keV energy band. The data were binned in ½ ACIS pixel bins and adaptively smoothed with scales from ½ to 20 image pixels (~0''.25 – 10'') in 30 iterations, with a minimum of 9 counts under the Gaussian kernel. The color bar at the bottom gives the number of counts per image pixel corresponding to each color.

Figure 2 shows an image of the central high surface brightness region in the 0.1-1 keV band, obtained with 1/16 pixel binning and adaptive smoothing (1 to 20 image pixels, with 30 iterations and a minimum of 5 counts under the kernel). Note the arc-like shape of the most intense emission in the south, which is likely to be the result of the pileup affecting the nuclear point source. Superimposed are the HST Hα+[NII] contours derived from Figure 2 of Rosario et al 2019. These contours were shifted by 0''.2 toward west to make them overlay onto the X-ray surface brightness. This shift is consistent with astrometric uncertainties between HST and *Chandra* that we have found in our previous work (Fabbiano et al 2018b). The shape of the X-ray surface brightness matches remarkably well with that of the Hα emission, justifying the shift.

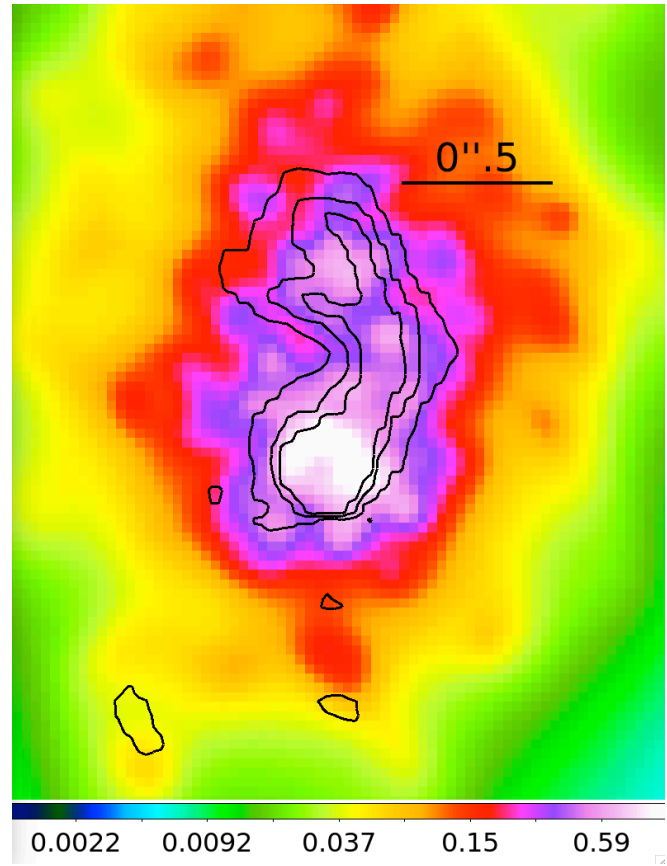

**Figure 2.** The central high surface brightness region in the 0.1-1 keV band, obtained with 1/16 pixel binning and adaptive smoothing (1 to 20 image pixels, with 30 iterations and a minimum of 5 counts under the kernel). Superimposed are the HST Hα+[NII] contours derived from Figure 2 of Rosario et al. (2019). The color bar at the bottom gives the number of counts per image pixel corresponding to each color.

To highlight the central high surface brightness emission, Figure 3 (left) shows the higher count-per-pixel central region, with 1/16 binning and a single 4-pixel Gaussian smoothing, with the Hα+[NII] contours. Figure 3 (right) also shows the counts detected in selected regions. The counts in the northern 'blob' are 37% of the counts in the emission associated with the nucleus to the south. Both this intensity and the position of this northern emission exclude that they may be due to the PSF anomaly[3]. The latter is expected to be a ~5% of the emission of the main source and would be positioned to the west of the nucleus in this exposure. The features we detect are highly statistically significant. Estimating the background emission from a 1''-2'' annulus surrounding both regions, we expect ~4 background counts in the northern region, ~6 counts in the nuclear region and ~2 counts in the box connecting the two.

Count pileup in ACIS-S affects the central count distribution of the nuclear point source (the main peak of peak at the south), depressing the central emission and producing an apparent arc-like feature, but cannot produce the kind of north-south distribution we detect. The read-out streak, visible at the higher energies, runs approximately in the east-west direction (see Figure 6 of Evans et al. 2006), so cannot affect the feature we detect.

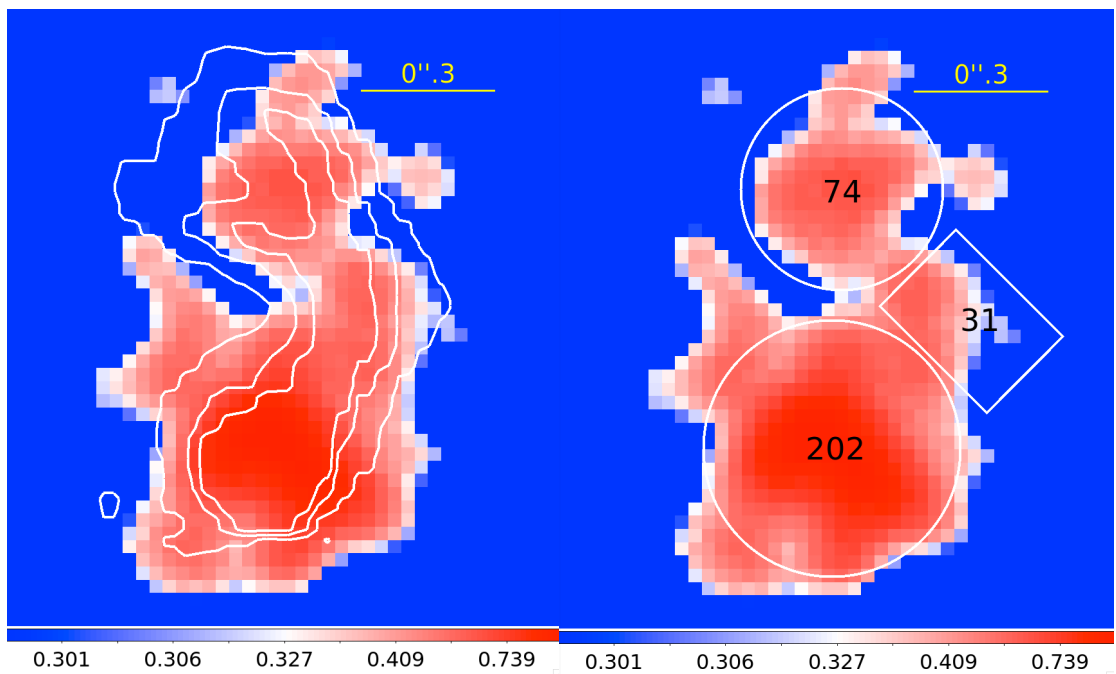

**Figure 3.** – Left: the higher count-per-pixel central region, with 1/16 binning and a single 4-pixel Gaussian smoothing, with the Hα+[NII] contours. Right: counts detected in selected regions. The color bar at the bottom gives the number of counts per image pixel corresponding to each color.

The spectrum of NGC 2110 shows a prominent Fe Kα line. Imaging in this line (using the 6.2-6.8 keV band) shows a single point source at the location of the southern more prominent 0.1-1.0 keV source in figures 2 and 3. This reinforces our conclusion that this southern source is at the location of the AGN.

---

[3] http://cxc.harvard.edu/ciao/caveats/psf_artifact.html

## 4. Discussion and Conclusions

Figure 4 shows the central ~5" of the soft (0.1 -1.0 keV) X-ray surface brightness distribution of NGC 2110, compared with the HST Hα+[NII] contours (black; derived from Figure 2 of Rosario et al. 2019), and the 5 GHz (VLA) contours from Evans et al. (2006), in white. While no obvious correspondence with the radio lobes can be found, except for a general N-S elongation, the areas of highest X-ray surface brightness are remarkably similar to the optical emission line distribution within 1" (170 pc) from the nucleus (see also Figures 2 and 3). This region is also consistent with that of most intense $H_2$ nuclear surface brightness, associated with a nuclear outflow (see Figure 3 of Diniz et al. 2015), and lies in the nuclear CO 2-1 'lacuna' discovered by Rosario et al. (2019), the dashed polygon in Figure 4.

These results suggest that X-ray emission can be quite effective in warming the molecular clouds and suppressing CO 2-1 emission. The results on NGC 2110 presented here are consistent with the conclusions emerging from our parallel ALMA work on the Compton thick AGN ESO 428-G014 (Feruglio et al. 2019). As reported here for NGC 2110, in ESO 428-G014 the most intense near-nuclear extended X-ray emission in the central few 100 pc is also coincident with excess $H_2$ emission and is anti-correlated with CO 2-1.

The ALMA data (Rosario et al. 2019) of NGC 2110 also show 1mm continuum within 2" of the nucleus, suggesting the presence of dense clouds, which is confirmed by the $H_2$ emission (Diniz et al 2015). In ESO 428-G014 the extended hard continuum (~3-6 keV) and Fe Kα emission (Fabbiano et al 2017; 2018a, b; 2019) require the presence of reflecting dense clouds, consistent with the ALMA detection of 1mm emission (Feruglio et al. 2019). If NGC 2110 is similar in its properties to ESO 428-G014, we would expect substantial extended hard (> 3 keV) X-ray emission in the circumnuclear region, where the ALMA result show that dense molecular clouds are clearly present (Rosario et al. 2019). The bright nuclear source of NGC 2110 impedes probing directly via imaging the presence of extended hard X-ray emission, but a suggestion of hard continuum emission at larger radii can be found in the analysis of Evans et al. (2006; see their Figure 7).

NGC 2110 provides a contrasting view of the AGN phenomenon to that shown by the ALMA and *Chandra* observations of the CT AGN NGC 5643 (Alonso-Herrero et al. 2018; Fabbiano et al. 2018c). In that case, a remarkable similarity is provided by the images of the CO 2-1 emission of a rotating circum-nuclear disk and Fe Kα emission of the same circum-nuclear emission. In these shielded regions, dense cold molecular clouds are detected directly and via the fluorescent emission of scattered X-ray photons.

The emerging picture is that of a multiphase ISM near the AGN, where less dense clouds are photoionized by the AGN, causing optical narrow line and soft X-ray emission, while denser molecular clouds may scatter the harder X-ray photons and also may be excited by the X-rays, as suggested by the NGC 2110 results. In the standard torus model of AGNs, these clouds would lie in the polar direction of the torus, along which the radio jets are also expanding. While the results of this paper prove the existence of X-ray emission in the excited warm molecular clouds, we cannot exclude that shocks may also at least in part contribute to this process. These clouds could also experience shocks, either because of nuclear winds, or interaction with the radio jets, as suggested by *Chandra* imaging/spectral observations of several nearby AGNs, including NGC 4151 (Wang et al. 2011a, b, c), Mkn 573 (Paggi et al. 2012) and ESO 428+G014 (Fabbiano et al. 2017, 2018a, 2018b, 2019); see also Evans et al. (2006).

This work demonstrates the importance of joint high-angular-resolution ALMA, *Chandra*, HST and IFU observations to probe the properties of the circumnuclear regions of the AGN and its interaction with the host galaxy. The results presented here argue for the need for the astronomical community to maintain this powerful multi-wavelength observational coverage in the years to come.

We retrieved data from the NASA-IPAC Extragalactic Database (NED) and the *Chandra* Data Archive. For the data analysis, we used the CIAO toolbox and DS9, developed by the Chandra X-ray Center (CXC). This work was partially supported by the NASA contract NAS8-03060 (CXC).

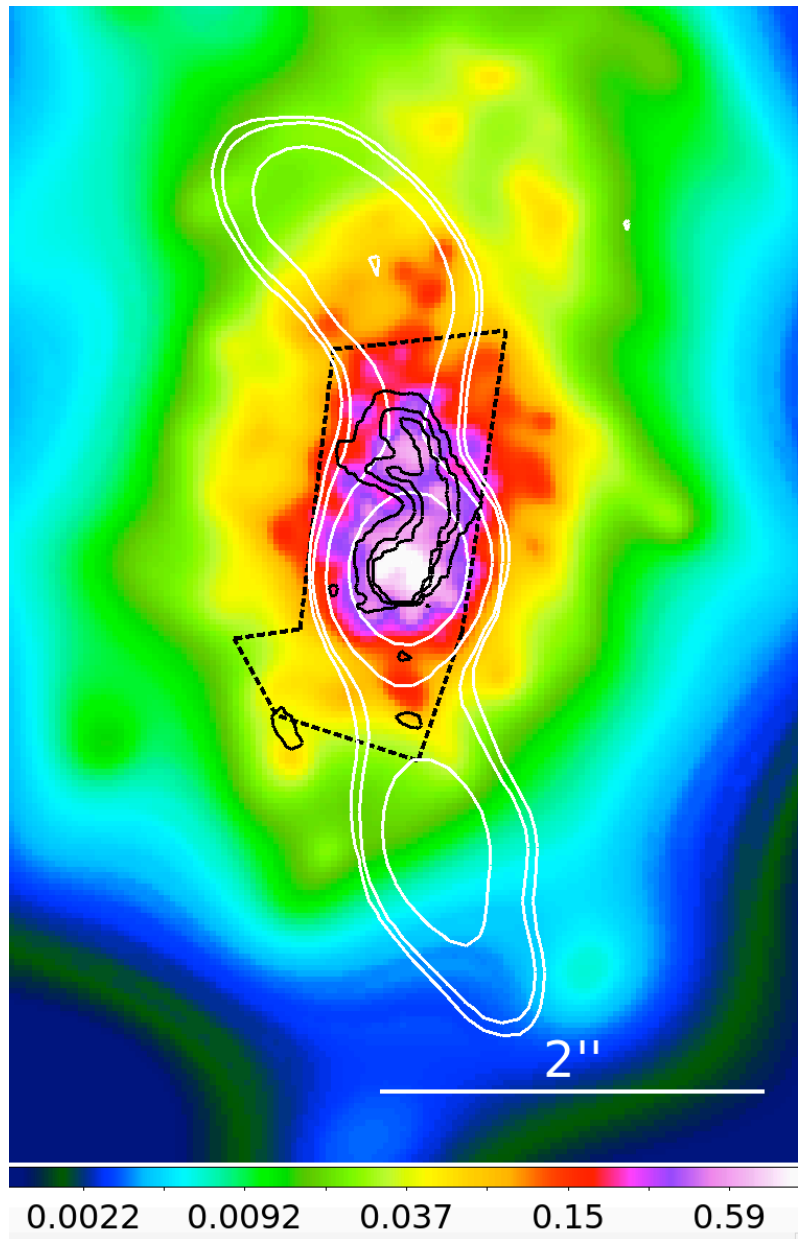

Figure 4. - The central high surface brightness region in the 0.1-1 keV band, obtained with 1/16 pixel binning and adaptive smoothing (1 to 20 image pixels, with 30 iterations and a minimum of 5 counts under the kernel). Superimposed are the HST Hα+[NII] contours (black) derived from Figure 2 of Rosario et al. (2019), a sketch of the CO 2-1 'lacuna' from Rosario et al. (2019), and the 5 GHz (VLA) contours from Evans et al. (2006), in white. The color bar at the bottom gives the number of counts per image pixel corresponding to each color.